\documentclass[conference]{IEEEtran}

\usepackage{comment}
\usepackage{physics}

\usepackage[table]{xcolor}
\usepackage{colortbl}
\usepackage{color}
\usepackage{float}
\definecolor{light_grey}{rgb}{0.8,0.8,0.8}
\usepackage{multirow}
\usepackage{diagbox}

\usepackage{algorithm}
\usepackage{algorithmic}
\usepackage{etoolbox}\AtBeginEnvironment{algorithmic}{\small}

\usepackage{graphicx}
\ifCLASSINFOpdf
\else
\fi
\usepackage{amsmath}

\usepackage{array}
\newcolumntype{L}{>{\centering\arraybackslash}m{3cm}}

\begin{document}

\title{Resource allocation in a Quantum Key Distribution Network with LEO and GEO trusted-repeaters \vspace{-3.5mm}}

\author{\IEEEauthorblockN{Milo Grillo\IEEEauthorrefmark{1}, Alexis A. Dowhuszko\IEEEauthorrefmark{2}, Mohammad-Ali Khalighi\IEEEauthorrefmark{3}, and Jyri Hämäläinen\IEEEauthorrefmark{2}} 

\IEEEauthorblockA{ \IEEEauthorrefmark{1}Department of Mathematics, ETH Z\"urich, 8092 Zurich, Switzerland\\
\IEEEauthorrefmark{2}Department of Communications and Networking, Aalto University, 02150 Espoo, Finland\\
\IEEEauthorrefmark{3}Aix-Marseille University, CNRS, Centrale Marseille, Institut Fresnel, Marseille, France\\
{ Email: milo.grillo@math.ethz.ch;  alexis.dowhuszko@aalto.fi;  ali.khalighi@fresnel.fr; jyri.hamalainen@aalto.fi}} \vspace{-2.5mm} } 

\maketitle

\begin{abstract}
Quantum Key Distribution~(QKD) is a technology that enables the exchange of private encryption keys between two legitimate parties, using protocols that involve quantum mechanics principles. The rate at which secret keys can be exchanged depends on the attenuation that is experienced. Therefore, it is more convenient to replace many terrestrial fiber segments (and repeaters) by just few optical satellite links that would enable flexible global coverage. Then, the satellite nodes can take the role of trusted-relays, forwarding the secret keys from source to destination. However, since the rate at which secret keys can be generated in each quantum link is limited, it is very important to select the intermediate satellite nodes to inter-connect ground stations efficiently. This paper studies the most convenient allocation of resources in a QKD network that combines complementary connectivity services of GEO and LEO satellites. The aim of the centralized routing algorithm is to select the most convenient trusted-relays to forward the secret keys between pairs of ground stations, verifying the constraints that satellite-to-ground and inter-satellite quantum channels have.  \end{abstract}

\vspace{2mm}


%

\vspace{-2mm}
\section{Introduction}
\label{sec:1}
\vspace{-0.5mm}

Quantum Key Distribution~(QKD) networks enable the distribution of secret keys between two legitimate parties by encoding the information in one of two randomly chosen non-orthogonal quantum states~\cite{benn2014}. The security of QKD is not based on the computational hardness of solving mathematical problems, but rather on physical processes that are not vulnerable to powerful computers~\cite{basc2013}. QKD can be also classified as an optical technology, which automates the delivery of encryption keys between any two points that share an optical link that could be either wired (fibers) or wireless (Free Space Optics). Unfortunately, QKD networks based solely on optical fibers face serious problems when trying to distribute secret keys in wide coverage areas. This is because the power loss that the physical communication channel introduces grows exponentially with distance, limiting the rate at which secret keys can be successfully exchanged over long coverage ranges~\cite{bedi2017}. 


The intrinsic \mbox{point-to-point} nature of a QKD system is a bottleneck for its applicability in global scale. Fortunately, the coverage range of a QKD system can be extended by using trusted-relays, which can be conveniently placed on satellite payloads to make them difficult to eavesdrop~\cite{simo2017}. Apart from providing better security, optical satellite links experience less attenuation than optical fiber links, as most of the propagation losses are concentrated in the low-layers of the atmosphere~\cite{toma2011}. Different satellite orbits can be used for this purpose, such as Geostationary~(GEO) and Low Earth Orbit~(LEO) satellites~\cite{basc2013}. A GEO quantum satellite can provide a slow but continuous secret key generation rate, due to its fixed position on the sky at a very high altitude. In contrast, LEO quantum satellites are much closer to the Earth's surface and, due to that, they can provide a faster but intermittent secret key generation service~\cite{huan2020}.

Most of the research done so far in the literature considered the use of LEO satellites for QKD, taking advantage of their low channel loss~\cite{bedi2017}. However, since a quantum LEO satellite is only visible to a particular Ground Station~(GS) for a limited time period, the secret key rate that is predicted is only available during the flyover time, few times a day~\cite{bour2013, liao2019}. Trying to provide a continuous QKD service, there are other authors that considered the use of a constellation of LEO satellites with inter-satellite links, similar to the IRIDIUM satellite system~\cite{prat1999},
or the addition of GEO satellites to enable continuous service~\cite{toma2011}. However, in most of these cases, the most convenient allocation of resources was not studied in detail for the whole (global) QKD network, or the routing and key allocation for the satellite-to-ground and inter-satellite links was done using heuristic algorithms that do not necessarily reflect the actual constraints that exist~\cite{huan2020}.




In this paper, we study the resource allocation problem of a QKD network that combines both GEO and LEO satellite constellations to enable an exchange of secret keys in a global coverage. The BB84 protocol with decoy state is considered in the quantum channels~\cite{lo2015}, and the relay of secret keys between ground stations is performed with the aid of trusted-relays in the satellites~\cite{liao2019}. By abstracting the GS and quantum satellites as \emph{nodes}, the quantum channels as \emph{edges}, and the amount of available secure keys as \emph{weights}, the satellite QKD network was modeled as a time varying \emph{graph}~\cite{li2020}. Then, the optimal routing and resource allocation for the QKD network is determined in a centralized way, solving an equivalent Linear-Programming problem with constraints in the GEO-to-GS, LEO-to-GS, and LEO-to-LEO quantum links.



The rest of the paper is organized as follows: Section~\ref{sec:2} presents the system model of the satellite QKD network, including the formulas that estimate rate at which secret keys can be generated for the space-to-ground and inter-satellite links. Section~\ref{sec:3} introduces the graph representation of the QKD network at a given time instant, and derives the algorithm that optimizes the routing and flow of secret keys in a centralized way. The parameters of the simulation setting, as well as the performance analysis of the obtained results, are presented in Section~\ref{sec:4}. Finally, conclusions and suggestions for future work are given in Section~\ref{sec:5}.

\vspace{-2mm}
\section{System model}
\label{sec:2}
\vspace{-0.5mm}

\begin{figure}
    \centering
    \includegraphics[width=0.48\textwidth]{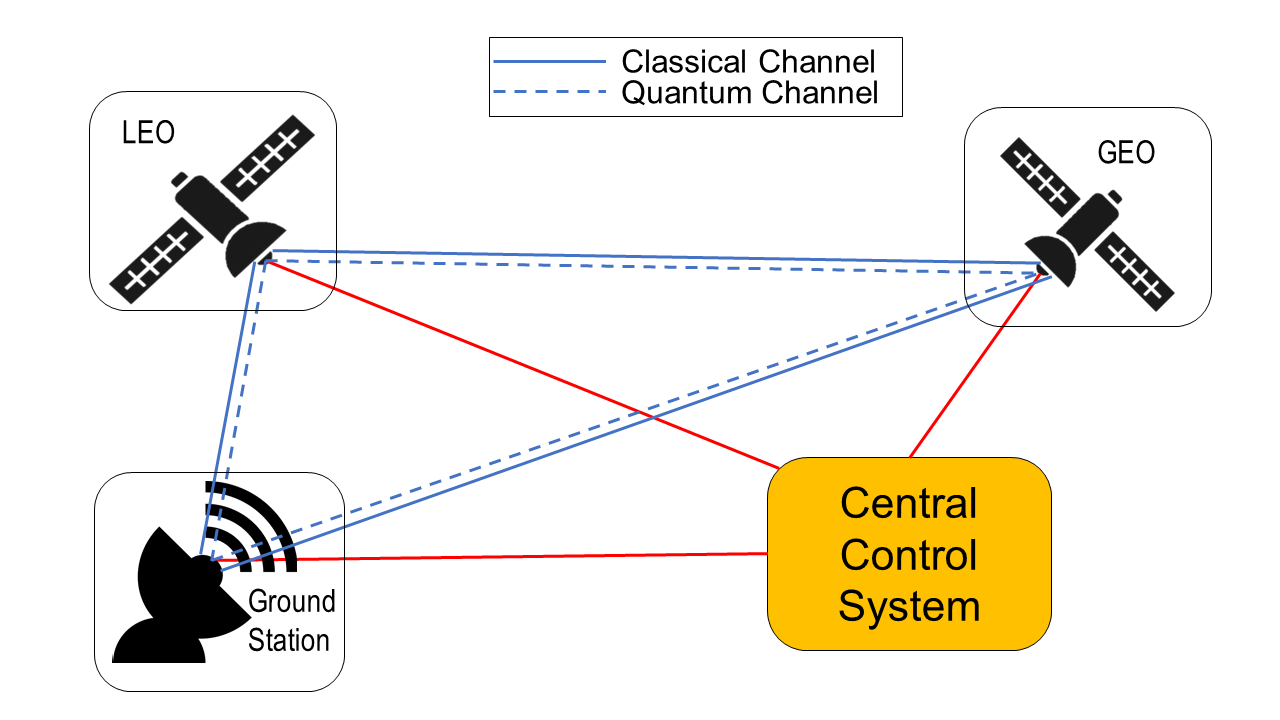}
    \vspace{-3mm}
    \caption{Overview of the QKD network that combines GEO and LEO trusted-repeaters. The Central Control Unit knows the rate at which secret keys can be generated on the ground-to-space and inter-satellite quantum channels (dashed blue lines), determines the most convenient routing on the classical channels (solid blue lines) to optimize the flow of secret keys, and informs these decisions to the QKD nodes using the control channels (red solid lines).}
    \label{fig:2.1}
    \vspace{-2mm}
\end{figure}

The simplified model of the satellite QKD network that combines both LEO and GEO trusted-repeaters is illustrated in Fig.~\ref{fig:2.1}. It consists of a constellation of (a few) GEO and (a number of) LEO satellites that provide service to a large number of GSs, which are sparsely deployed on the globe. Similar to terrestrial QKD networks based on optical fibers, the nodes of the satellite QKD network have quantum communication channels that are used to generate secret keys, and classical communication channels that are used to transport the QKD protocol signaling as well as to forward encrypted secret keys between non-directly connected nodes.   

Let us also assume that the service area of the whole QKD network is divided into non-overlapping regions served by a GS, whose associated users receive the secret keys using terrestrial optical fiber networks. Without loss of generality, we consider that each GS is equipped with three Free-Space Optical~(FSO) transceivers~\cite{khal2014}, which enable the connectivity with the GEO satellite and (up to two) LEO satellites that may be visible on the sky at each time window. Each LEO satellite relies on two FSO links for inter-satellite connectivity and two FSO links for space-to-ground QKD (\emph{i.e.}, LEO-to-GS). Note that the GEO satellite is always visible to the same GSs, whereas the visibility of a LEO satellite towards a given GS varies with the time of the day. Finally, it is considered that the centralized resource manager has access to the status of the whole QKD network in each time window (\emph{i.e.}, secret key pools in nodes and secret key rates per FSO link), and that is able to determine the most convenient route to exchange secret keys between each given pair of remote GSs.


\subsection{Decoy state Quantum Key Distribution}
\label{sec:2a}


In this paper, the QKD between space nodes (GEO/LEO) and ground nodes (GS) is considered to be carried out in downlink, from space-to-ground. The QKD transmitters in the satellites use weak coherent laser pulses to implement the decoy-state Bennett-Brassard 1984~(BB84) protocol~\cite{benn2014}, which is immune to photon-number-splitting attacks.

Let us assume that we implement the BB84 protocol with vacuum-plus-weak-decoy-state~\cite{lo2015}. That is, we consider that Alice can prepare and emit a weak coherent state $\ket{\sqrt{\mu} \, {\rm e}^{{\rm j} \theta}}$. Assuming that phase $\theta$ of each signal is randomized, then the number of photons of the signal state follows a Poisson distribution with parameter $\mu$, which represents the intensity of the signal states. Here, 
\begin{equation}
P_n(\mu) = \big( \mu^n \, {\rm e} ^{- \mu} \big)/ n!,
    \label{eq:2.1}
\end{equation}
is the probability that the pulse generated by Alice contains $n$ photons. So, it is assumed that any mixture of photon number states with Poisson distribution can be prepared by Alice, with an intensity that can be changed for each individual pulse.

A lower bound for the rate at which secret keys can be generated in this situation was presented in~\cite{lo2015}, \emph{i.e.},
\begin{equation}
R_{\rm bb84} \ge q \, \Big\{ - Q_{\mu} \, f(E_{\mu}) \, H_2(E_{\mu}) + Q_1 \big[ 1 - H_2 (e_1) \big] \Big\}, 
    \label{eq:2.2}
\end{equation}
where $Q_{\mu}$ ($Q_{\nu}$) and $E_{\mu}$ ($E_{\nu}$) are the gain and Quantum Bit Error Rate~(QBER) of the signal (weak decoy) states~$\mu$~($\nu$), whereas $Q_1$ and $e_1$ are the gain and the error rate of \mbox{single-photon} states. Moreover, $q=1/2$ is the efficiency of the BB84 protocol, $f(x) = 1.22$ is the bi-directional error correction efficiency for the Cascade protocol, and
\begin{equation}
H_2(x) = -x \, \log_{\rm 2}(x) - (1 - x) \log_{\rm 2} (1 - x)
\label{eq:2.3}
\end{equation}
is the binary Shannon entropy. For a coherent state, the gain and QBER of the signal states is given by
\begin{equation}
Q_{\mu} = \sum_{n=0}^{\infty} Y_n \, P_n(\mu), \quad E_{\mu} = \bigg( \sum_{n=0}^{\infty} Y_n \, P_n(\mu) \, e_n \bigg) / Q_{\mu},
\label{eq:2.4}
\end{equation}
where
\begin{equation}
    Y_n = Y_0 + \delta_n - Y_0 \, \delta_n \approx Y_0 + \delta_n
    \label{eq:2.5}
\end{equation}
is the probability that Bob's measurement is conclusive when Alice emits an $n$-photon pulse, and
\begin{equation}
e_n = Y_0 / (2 \, Y_n), \qquad \delta_n = 1 - (1 - \delta)^n
    \label{eq:2.6}
\end{equation}
are the error rate and attenuation of the $n$-photon signals.

The values of $Q_{\mu}$ and $E_{\mu}$ can be estimated directly from~\eqref{eq:2.4}. However, the value of $Q_1$ cannot be determined in closed form and needs to be bounded based on other gains. According to~\cite{ma2005}, the gain and error rate of the single-photon states when using the vacuum-plus-weak-decoy-states method verify the following lower~(L) and upper~(U) bounds, respectively: 
\begin{equation}
Q_1^{\rm L} = \mu \, {\rm e}^{-\mu} \, Y_1^{\rm L} \le Q_1, \qquad e_1^{\rm U} = \frac{E_{\nu} \, Q_{\nu} \, {\rm e}^{\nu} - e_0 Y_0}{Y_1^{\rm L} \, \nu},
\label{eq:2.7}
\end{equation}
where $E_{\nu}$ and $Q_{\nu}$ are obtained replacing $\mu$ for $\nu$ in~\eqref{eq:2.4}, and
\begin{equation}
Y_1^{\rm L} = \frac{\mu}{\mu \nu - \nu^2} \bigg( Q_{\nu} \, {\rm e}^{\nu} - Q_{\mu} \, {\rm e}^{\mu} \frac{\nu^2}{\mu^2} - \frac{\mu^2 - \nu^2}{\mu^2} Y_0 \bigg) \le Y_1.  \label{eq:2.8}       
\end{equation}

The background yield $Y_0$ can be computed as the gain of the vacuum decoy state, whereas the background error rate $e_0 = 1/2$ due to dark counts happen randomly, so half of the times photons click on the correct detector in this situation. 

\vspace{-2mm}
\subsection{Satellite based QKD network based on trusted-repeaters}
\label{sec:2b}
\vspace{-0.5mm}


Quantum satellites can be used as trusted-repeaters to generate secret keys between distant nodes that do not share a QKD link in common. For example, let us assume that source node~$S$ wants to generate a secret key with destination node~$D$. To achieve this goal, $S$ starts the process by generating a random key $K_{{\rm s},{\rm d}}$ of suitable length, and forwards it to an intermediate node $I_1$ performing a bit-wise exclusive OR~(XOR) operation~(denoted by $\oplus$) with a secret key $K_{{\rm s},{\rm i_1}}$ of the same length that shares with  $I_1$. This new string ($K_{{\rm s},{\rm i_1}}\oplus K_{{\rm s},{\rm d}}$) can be sent through a classical communication channel from $S$ to $I_1$, who can decode the original key with another XOR operation (\emph{i.e.}, $ K_{{\rm s},{\rm d}} = ( K_{{\rm s},{\rm i_1}} \oplus K_{{\rm s},{\rm d}}) \oplus K_{{\rm s},{\rm I_1}}$. Then, the XOR operation is performed with the secret key $K_{{\rm I_1},{\rm I_2}}$ shared between node $I_1$ and $I_2$, the resulting string of bits is transmitted over a classical communication channel, and the original key $K_{{\rm s},{\rm d}}$ is recovered in the intermediate node $I_2$. Note that this procedure is repeated until the secret key $K_{{\rm s},{\rm d}}$ reaches $D$, consuming the same amount of secret keys in each of the links that are used. 

The XOR operation that is performed to forward the random keys from the source GS to the destination GS consumes the secret keys that are available in each link. For this purpose, each LEO-to-GS, GEO-to-GS, and LEO-to-LEO link must generate secret keys continuously, and store them in quantum key pools associated to the different links of the QKD network. Due to the rate at which secret keys can be generated is limited, the centralized control system should optimize the route selection and key assignment, such that the limited resources of the QKD network are efficiently used. 

Without loss of generality, we assume that the centralized controller has continuous access to the status of the whole QKD network, and is able to perform the resource allocation to optimize the target objective, such as the maximization of key flows and the minimization of secret key consumption.

\vspace{-2mm}
\subsection{Attenuation of space-to-ground and inter-satellite links}
\label{sec:2c}
\vspace{-0.5mm}


The total attenuation that an FSO link experiences is defined as the ratio between the mean transmit and receive powers, measured at the entrance and exit of the transmit and receive telescopes. When the optical receiver is placed in the far field of the transmitter (\emph{i.e.}, when the link distance $L \ge D_{\rm T}^2 / \lambda$), the attenuation due to diffraction in the FSO link is given by~\cite{pfen2003}
\begin{equation}
    \delta_{\rm diff} = \frac{L^2 \Big( \theta_{\rm T}^2 + \theta_{\rm atm}^2 \Big)}{D_{\rm R}^2} \frac{1}{T_{\rm T} \Big( 1 - L_{\rm P}\Big) T_{\rm R} } , 
    \label{eq:2.9}  
\end{equation}
where $\lambda$ is the wavelength, $D_{\rm T}$ ($D_{\rm R}$) is the diameter of the transmit (receive) telescope, $T_{\rm T}$ ($T_{\rm R}$) is the transmit (receive) telescope transmission factor, and $L_{\rm P}$ is the pointing loss due to misalignment between the transmitter and the receiver. 
The divergence angle resulting from the transmit telescope can be approximated by $\theta_{\rm T} = \lambda / D_{\rm T}$, and the additional divergence caused by the atmospheric turbulence is given by $\theta_{\rm atm} = \lambda / r_0$, where $r_0$ is the Fried parameter.  

The attenuation due to diffraction is originated by the natural beam-broadering that light experiences when propagating, which makes that a certain faction of the transmitted power cannot be collected at the receiver when the received beam spot diameter is larger than the aperture of the receive telescope. Apart from the geometric loss in~\eqref{eq:2.9}, there are other losses to be considered when estimating the rate at which secret keys can be generated in a quantum channel. The total attenuation that the optical wireless link experiences is \begin{equation}
    \delta = \delta_{\rm diff} \times \delta_{\rm atm} \times \delta_{\rm rec}, \quad \delta_{\rm atm} = \delta_{\rm abs} \times \delta_{\rm scat} \times \delta_{\rm turb},   
\end{equation}
where $\delta_{\rm rec}$ is the loss due to inefficiencies in the photon detection process and $\delta_{\rm atm}$ is the attenuation in the atmosphere originated in the absorption ($\delta_{\rm abs}$) and scattering ($\delta_{\rm scat}$) imposed by the constituent gases and particles of the atmosphere, as well as the atmospheric turbulence ($\delta_{\rm turb}$) caused by random fluctuations in the refractive index of the light-beam path.  


\vspace{-2mm}
\section{Centralized optimization of the hybrid satellite QKD network}
\label{sec:3}
\vspace{-0.5mm}

In this section, we introduce the  graph representation of the QKD network topology and derive the algorithms that optimize the routing and secret key flows in a centralized way.
 
 \vspace{-2mm}
\subsection{Dynamic Graph representation}
\label{sec:3a}
\vspace{-0.5mm}

All transceivers in our QKD network are represented by nodes in a weighted temporal graph. The weight of the edges between two nodes represents the secret key transfer rate between them. In our satellite QKD network, we distinguish between three types of nodes connected by FSO links, namely:
\begin{itemize}
    \item \emph{GS nodes}, which are fixed on the Earth and can communicate with few GEO/LEO satellites at the same time. A GS aims at exchanging secret keys with other GSs, with whom it does not have a direct quantum communication channel. In graph technical terms, GS acts as either a source/destination node or as trusted-relay.
    \item \textit{GEO satellite}, which is stationary in an orbit that is relatively far from Earth's surface (\emph{i.e.}, at about $36000$\,km). Due to that, the rate at which secret keys can be generated in GEO-to-GS links is relatively slow but constant. In the graph, a GEO node can only act as a relaying node.
    \item \textit{LEO satellites}, which are not stationary with respect to the Earth as they usually move from pole-to-pole, in sun-synchronous orbits. As a result, the rate at which secret keys can be generated varies in LEO-to-GS links, but may remain constant between LEOs in the same orbit. The secret key rate is maximal when the distance between nodes is minimal. A LEO can only act as a relaying node.
\end{itemize}

\begin{figure}[t]
    \centering
    \includegraphics[width=0.5\textwidth]{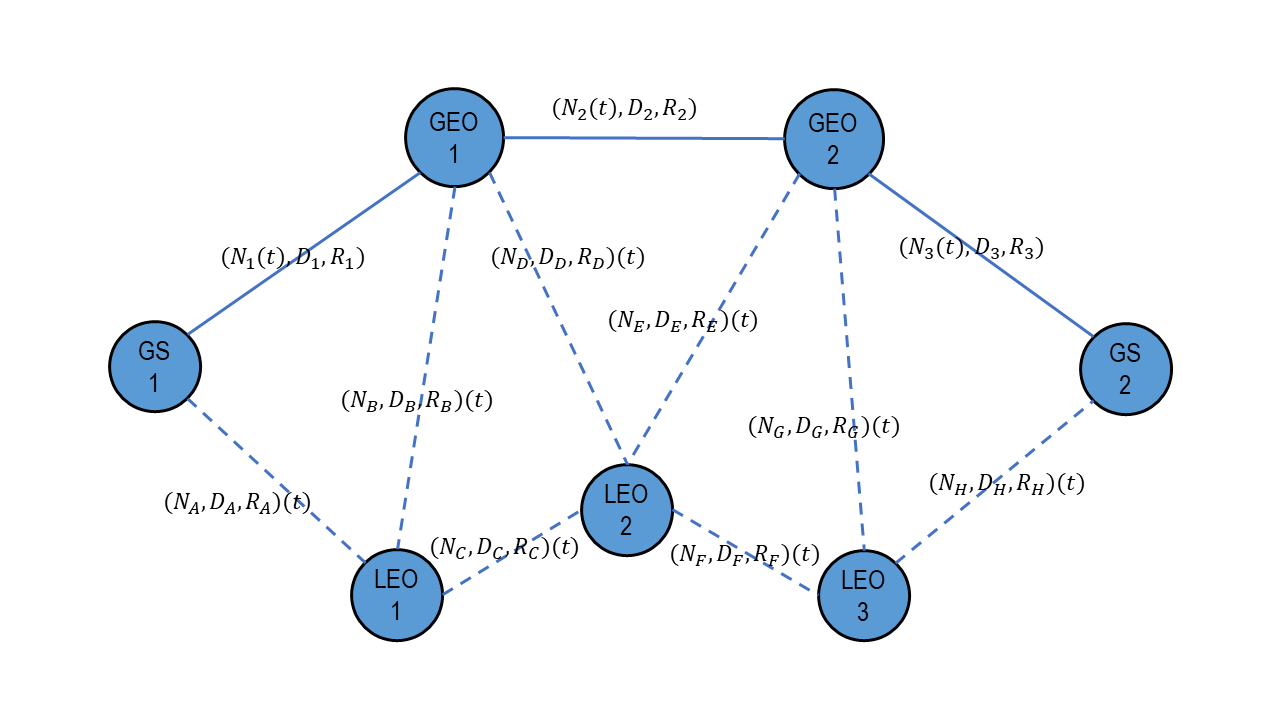}
    \vspace{-10mm}
    \caption{Graph representation of a simplified QKD network topology that combines two GEO satellites, three LEO satellites, and two ground stations.}
    \label{fig: example_graph}
    \vspace{-5mm}
\end{figure}

A very simple example of this graph representation is given in Fig.~\ref{fig: example_graph}. The time-dependent connections are represented using dotted lines. Note that the amount of keys in the quantum key pool is, strictly speaking, time-dependent as the keys on a given link may be generated and consumed at variable rates by the trusted-relays. The generation rate $R_i$ depends on the distance~$D_i$ between the nodes, but also on atmospheric losses for space-to-ground links (\emph{e.g.} in case of clouds or clear skies). 

\vspace{-2mm}
\subsection{Multi-commodity flow problem}
\label{sec:3b}
\vspace{-0.5mm}

Consider a graph similar to the one shown in Fig.~\ref{fig: example_graph}, with $N_{\rm gs}$ ground stations, all connected by a network of LEO and GEO satellites. If we simply want to find the maximum amount of secret keys that GS~$A$ can exchang with GS~$Z$, we can formulate this optimization problem as a \textit{max-flow problem} and continue to solve it using straightforward path-finding techniques. When multiple GSs want to exchange secret keys with multiple other GSs, one may be tempted to formulate the optimization problem as a multi-source, multi-sink max-flow problem. Although this formulation may be straightforward, it fails to consider that the keys sent out by a given source GS are aimed towards a specific destination GS, not just any arbitrary GS open to accept secret keys (\emph{i.e.}, the keys from the different GS should be considered as unique \textit{commodities}). 

This fact leads us to the \textit{multi-commodity flow problem} formulation. Let us consider a flow network $G = (V, E)$, where each edge $(v, w) \in E$ has a maximum capacity $u(v, w) \geq 0$. In our system, the capacity is the amount of secret keys in the key pool associated to the link $(v, w)$. In our graph, $k$ of the $N$ GSs wish to exchange secret keys with other $k$ GSs. So, we consider $k$ commodities $K_1, K_2, \dots, K_k$ with $K_i = (s_i, t_i, d_i)$, where $s_i$, $t_i$, and $d_i$ are the standard source, sink, and \textit{demand} of commodity $K_i$, respectively. We define $f_i(v, w)$ as the flow of commodity $K_i$ on edge $(v, w)$, which is the amount of secret keys of sort $K_i$ that are sent over the link $(v, w)$. This flow has a few natural restrictions:
\begin{enumerate}
    \item[A)] \emph{Link capacity}. It is not possible to send more secret keys than the maximum amount dictated by the link capacity. That is, $\sum_{i \in K} f_i(v, w) \leq u(v, w)$, $\forall (v, w) \in E$.
    \item[B)] \emph{Positive flow.} It is not possible to send a negative amount of secret keys, \emph{i.e.}, $f_i(v, w) \geq 0$, $\forall (v, w) \in E, i \in [k]$.
    \item[C)] \emph{Flow conservation.} At each relaying node, the same amount that flows in should flow out. However, for the source and sink nodes, the flow should be (up to a minus sign) equal to the demand. That is, $\forall v \in V, i \in [k]$, \vspace{-2mm}
    \begin{equation}
        \hspace{-1.0mm} \sum_{w \in N_v} \hspace{-1.5mm} f_i(v, w) \hspace{-0.5mm} - \hspace{-2.5mm} \sum_{w \in N_v} \hspace{-1.5mm} f_i(w, v) \hspace{-0.5mm} = \hspace{-0.5mm} \begin{cases} 0 &\text{if $s_i \hspace{-0.5mm} \neq \hspace{-0.5mm} v \hspace{-0.5mm} \neq \hspace{-0.5mm} t_i$}  \\ d_i &\text{if $v=s_i$} \\ -d_i &\text{if $v=t_i$} \end{cases} \hspace{-0.5mm}
     \vspace{-1mm}
    \end{equation}
    where $N_v$ is the number of neighboring nodes of $v$.
\end{enumerate}

A flow graph with such restrictions allows for several possible optimizations. One option is to maximize the total demand, i.e. a formulation in which $d_i$'s are not fixed and the goal is to maximise $\sum_{i \in [k]} d_i$. Similarly, one could try to maximise $\min_{i \in [k]}(d_i)$. A third optimization option could aim at, for a given a set of demands $\{d_i\}_{i \in [k]}$, minimize the total flow on the QKD network, \emph{i.e.} to consume the least amount keys in the pools. Additionally, each flow can receive a weight that represents the relative cost of using each FSO link. All these optimizations have different purposes and interpretations. In this paper, we focus on the later two options: on one hand, we consider the max-min demand optimization, which ensures that all GS pairs generate \emph{at least} a given amount of secret keys, such that the path of one request does not hinder another requests. On the other hand, we consider the min flow optimization, which ensures that the \emph{least} amount of keys is consumed to fulfill all requests. Note that the later approach is useful when the QKD network needs to be prepared to handle future requests of key generations. 

In case we only allow integer values for $f_i$, these optimization problems become NP-complex~\cite{even1975complexity}. Though it makes sense to have integer-valued flows in the QKD network (as keys cannot be split into parts), a relaxation that allows $f_i$ to take fractional values can be applied to solve the optimization problem using Linear Programming~(LP) schemes. Then, in favor of the reduced algorithmic complexity, fractional-valued solutions can be first found and then rounded down to an integer value. This is justified by the fact that the demand $d_i$ in QKD is typically on the order of hundreds or thousands; so, by rounding down, the relative loss would be negligible.

The round down processing guarantees as well that the resulting flow-paths are feasible. However, it may also result in an \emph{unused} link capacity. Therefore, after the first part of the algorithm is over, we obtain a new graph with reduced capacity. Then, we greedily choose the commodity $K_i$ with the lowest filled demand $d_i$, and find a path from source $s_i$ to sink $t_i$ such that exactly one key can be sent over this path. If a path is found, the demand $d_i$ is increased by one, and the found path is added to the solution. If no path is found, the algorithm ends. This procedure is summarized in Algorithm~\ref{alg: rounding}.

\begin{algorithm}[t]
    \caption{Greedy Rounding}
    \label{alg: rounding}
    \begin{algorithmic}[1]
        \REQUIRE A Graph $G = (V, E)$, with edge capacities, founds flows $f$ per edge and commodity and demands $d$ per commodity
        \STATE Round down flows in $f$ and adjust demands $d$ accordingly
        \STATE Subtract all the flows from their respective edge capacities
        \WHILE{available commodities exist}
            \STATE Pick $\hat k$ with lowest demand from the available commodities
            \STATE Using Dijkstra's algorithm, find the path which consumes least secret keys to send one additional key from source to sink
            \IF{a route exists}
                \STATE Add one to the demand and respective flows in found path; subtract one from the capacity of the edges on the path
            \ELSE
                \STATE Remove $\hat k$ from the available commodities
            \ENDIF
        \ENDWHILE
        \RETURN All rounded down flows and demands.
    \end{algorithmic}
    \vspace{-1mm}
\end{algorithm}

To find the optimal fractional flows, we use a LP algorithm, which aims at finding a vector $\mathbf{x}$, such that $\mathbf{c}^{\rm T} \mathbf{x}$ is minimized while verifying $\mathbf{A} \mathbf{x} \leq \mathbf{b}$, where $\mathbf{c}$ and $\mathbf{b}$ are two appropriate vectors and $\mathbf{A}$ is a matrix. The inequality is to be understood element-wise. By proper manipulation, the LP formulation can also allow equality restrictions. Standard LP problems can be solved using python packages, such as scipy~\cite{2020SciPy-NMeth}\footnote{Scipy's linprog optimization allows for upper bound constraints in the form $A_{\rm ub}x = b_{\rm ub}$, a matrix equality $A_{\rm eq}x = b_{\rm eq}$, and strict bounds $L \leq x \leq U$ which significantly simplifies the notation.}.

The restrictions A, B and C, as given above, allow for a LP formulation. By choosing the parameter $\mathbf{x}$ as the demands and the flows of all commodities, each edges in two directions filling the matrix $\mathbf{A}$ with $0$, $1$ and $-1$ at the correct places, and $\mathbf{b}$ with either $0$ or the edge capacities, one can write the (in)equalities from the considered restriction in matrix form. Note that the size of $\mbox{x}$ is $\#commodities \times (2 \times \#edges + 1)$ if we keep $d_i$ variable, and $\#commodities \times 2 \times \#edges$ if we fix $d_i, \forall i \in[k]$. The algorithm for max-min demand optimization and the minimum resource usage given a set of requests are summarized as Algorithms \ref{alg: max_min} and \ref{alg: min_resource}, respectively\footnote{In the algorithms, we note  that if $f(v, w) = f_j^k$, then $f(w, v) = \hat f_j^k$.}.  

\begin{algorithm}[t]
    \caption{Routing: Maximise minimum demand (MMD)}
    \label{alg: max_min}
    \begin{algorithmic}[1]
        \REQUIRE A Graph $G = (V, E)$ with capacities per edge
        \FORALL{GS pairs $(a, b)\in \{a, b: a,b \in V_{\rm gs}, a \neq b\}$}
            \STATE Create a commodity $K_i = (a, b, d_i)$, where $d_i$ is variable
        \ENDFOR
        \STATE Translate graph restrictions to LP matrix notation, \emph{i.e.}, find $\mathbf{A}$ and $\mathbf{b}$ such that $\mathbf{A} \, \mathbf{x} \leq \mathbf{b}$, where $\mathbf{x} = (t, d_1, \dots d_k, f_1^1, \hat f_1^1, \dots, \hat f_n^k)$, with $t$ a dummy variable and cost $\mathbf{c} = (-1, 0, \dots, 0)$
        \STATE Find $\mathbf{x}$ by LP
        \STATE Round down flows and demands in $\mathbf{x}$ using Algorithm~\ref{alg: rounding}
        \RETURN All values in $\mathbf{x}$ except $t$.
    \end{algorithmic}
\end{algorithm}

\begin{algorithm}[t]
    \caption{Routing: Minimize resource usage (MR)}
    \label{alg: min_resource}
    \begin{algorithmic}[1]
        \REQUIRE A Graph $G = (V, E)$ with capacities per edge. A set of key exchange requests $r$ with given amounts
        \FORALL{exchange request $r = (s_i, t_i, d_i)$}
            \STATE Create a corresponding commodity $K_i = (s_i, t_i, d_i)$
        \ENDFOR
        \STATE Translate graph restrictions to LP matrix notation, \emph{i.e.}, find matrix $\mathbf{A}$ and vector $\mathbf{b}$ such that $\mathbf{A} \, \mathbf{x} \leq \mathbf{b}$, where $\mathbf{x} = (f_1^1, \hat f_1^1, \dots, \hat f_n^k)$ and a relative cost $ \mathbf{c} \stackrel{\text{\tiny by default}}{=} (1, 1, \dots, 1)$
        \STATE Find $\mathbf{x}$ by LP
        \IF{no solution is found}
            \STATE End algorithm
        \ENDIF
        \STATE Round down flows and demands in $\mathbf{x}$ using Algorithm~\ref{alg: rounding}
        \RETURN All values in $\mathbf{x}$.
    \end{algorithmic}
\end{algorithm}

\section{Simulation results}
\label{sec:4}

This section presents the parameters of the simulation setting and the performance analysis of the proposed approach. 

\subsection{Parameters of the simulation scenario}
\label{sec:4a}

\begin{figure}[t]
    \centering
    \vspace{-4.5mm}
    \includegraphics[width=0.49\textwidth]{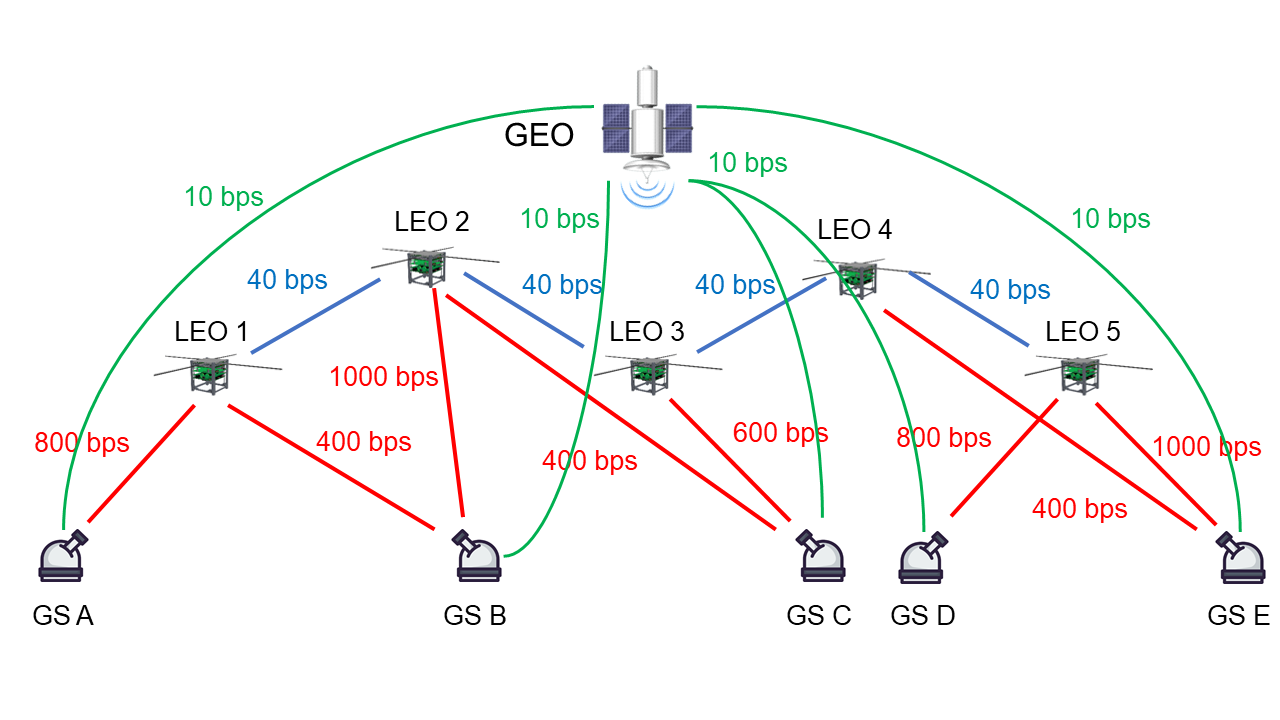}
    \vspace{-12mm}
    \caption{Schematic representation of the GSs and the satellite constellations setup for a given time window. Each edge specifies the key generation rate.}
    \label{fig: simulation_graph}
\end{figure}

The satellite constellation used in the simulations is shown in Fig.~\ref{fig: simulation_graph}, where on each edge, the secret key generation rate is specified in bits-per-second~(bps). In this setup, we assume that a GS may act as trusted-repeater if convenient. Note that if only one FSO link is enabled per GS, the option to make it act as trusted-repeater becomes unfeasible. Without loss of generality, we assume that QKD network in Fig.~\ref{fig: simulation_graph} has been up for exactly one minute, starting from empty quantum key pools. This gives a setup with reasonable ratios between the sizes of the quantum key pools of the links. We investigate the case where the goal is to maximise the minimum of the met demands (\emph{i.e.}, $ \max \min_i {d_i}$), as well as the situation in which a set amount of requests is given.

In satellite-based quantum communications, the uplink and downlink optical wireless channels are very different. Since the atmospheric turbulence occurs only in the last part of the downlink propagation path, near the terrestrial GS, the width of the light beam that enters the atmosphere is usually larger than the scale of the turbulent eddies. Due to that, the power loss due to the beam-wandering effect is minimal, and the attenuation losses are dominated by the diffraction effects considered in~\eqref{eq:2.9}. Scintillation can occur at some extend, but the averaging effect of large ground telescopes makes the effect of turbulence negligible~\cite{hoss2019}. The FSO link budget parameters for the different space-to-ground and inter-satellite links are summarized in Table~\ref{tab:4.1}. On the other hand, Table~\ref{tab:4.2}, gives the parameters of the BB84 protocol with weak decoy states that was used, and makes an estimation of the secret key rates generated on the different QKD network links.

\begin{table}[t]
\vspace{-3.5mm}
\centering
\caption{FSO link parameters for space-to-ground (LEO-to-GS and GEO-to-GS) and inter-satellite (LEO-to-LEO) links.}
\vspace{-2mm}
\label{tab:4.1}
\begin{tabular}{|c|c|c|c|}
 \hline
\rowcolor{light_grey} 
Parameter (Notation) &  \hspace{-1mm}
 LEO-to-GS \hspace{-1mm}
& \hspace{-1mm}
 GEO-to-GS \hspace{-1mm}
& \hspace{-1mm}
 LEO-to-LEO \hspace{-1mm}  \\
 \hline
Wavelength ($\lambda$) & $850$\,nm  & $650$\,nm & $1550$\,nm \\
\hline
\hspace{-5mm}
 Transmitter aperture ($D_{\rm T}$) \hspace{-5mm}
 & \multicolumn{3}{c|}{$30$\,cm } \\
\hline
Receiver aperture ($D_{\rm R}$) & $100$\,cm & $100$\,cm & $30$\,cm \\
\hline
Pointing loss ($L_{\rm P}$) & $7$\,dB & $1$\,dB & $3$\,dB \\
\hline
\hspace{-5mm}
 Telescope factors ($T_{\rm T}$\,/\,$T_{\rm R}$) \hspace{-5mm}
 & \multicolumn{3}{c|}{$0.8$\,/\,$0.8$  } \\
\hline
Detector efficiency ($\delta_{\rm rec})$ & $65$\% & $65$\% & $65$\% \\
\hline
Atmospheric loss ($\delta_{\rm atm})$ &  $1$\,dB & $1$\,dB &  $0$\,dB \\
\hline
\hspace{-3mm}
\multirow{2}{*}{Link range ($L_{\min}; L_{\max}$)} \hspace{-3mm}
&  ($800$\,km; & ($36000$\,km; &   \multirow{2}{*}{$4000$\,km} \\
 & $1200$\,km) & $42000$\,km) & \\
\hline
%
\end{tabular}
      \vspace{-5mm}
\end{table}

\begin{table}[t]
\centering
\caption{Parameters for the BB84 protocol with weak-decoy-states.}
\vspace{-2mm}
\label{tab:4.2}
\begin{tabular}{|c|c|c|c|}
 \hline
\rowcolor{light_grey} 
Parameter &  LEO-to-GS & GEO-to-GS & LEO-to-LEO  \\
 \hline
Wavelength ($\lambda$) & $850$\,nm  & $650$\,nm & $1550$\,nm \\
\hline
Link distance ($L$) & $1000$\,km  & $39000$\,km & $4000$\,km \\
\hline
QKD scheme & \multicolumn{3}{c|}{BB84 weak decoy states ($\mu = 0.3$; $\nu = 0.1$)  } \\
\hline
Dark counts ($Y_0$) & \hspace{-5mm} $1.7 \times 10^{-6} \hspace{-5mm} $ & $1.7 \times 10^{-6}$ & $1.7 \times 10^{-6}$
\\
\hline
Gain signal states ($Q_{\mu}$) & \hspace{-5mm}  $1.96 \times 10^{-3}$ \hspace{-5mm}  & $1.27 \times 10^{-5}$ & \hspace{-5mm}  $2.26 \times 10^{-5}$ \hspace{-5mm} 
\\
\hline
\hspace{-5mm}  QBER signal states\,($E_{\mu}$) \hspace{-5mm}   & $0.04$\% & $6.68$\% & $3.76$\%
\\
\hline
\hspace{-5mm} Gain vacuum states\, ($Q_{\nu}$) \hspace{-5mm} & \hspace{-5mm}  $3.28 \times 10^{-4}$ \hspace{-5mm}  & $5.38 \times 10^{-6}$ & \hspace{-5mm}  $8.66 \times 10^{-6}$ \hspace{-5mm} 
\\
\hline
\hspace{-5mm} QBER vacuum states\,($E_{\nu}$) \hspace{-5mm} & $0.26$\% & $15.81$\% & $9.81$\%
\\
\hline
\hspace{-5mm} Secret key rate\,($R_{\rm bb84}$) \hspace{-5mm}  & \hspace{-5mm}   $\sim  1000$\,bps \hspace{-5mm}  & $\sim 10$\,bps & $\sim 40$\,bps
\\
\hline
Pulse repetition rate & \multicolumn{3}{c|}{$10 \times 10^6$ pulses/sec.  } \\ 
\hline
\end{tabular}
      \vspace{-3mm}
\end{table}

\vspace{-2mm} 
\subsection{Performance analysis}
\label{sec:4b}
\vspace{-0.5mm} 

In very simple cases, only the isolated exchange of secret keys between two GSs is required. Consider \emph{e.g.} the exchange of secret keys between GS~$A$ and GS~$B$. Table~\ref{tab: simple_comparison} compares the multi-commodity flow method with other methods, such as Dijkstra that simply chooses the shortest available path. The first row of Table~\ref{tab: simple_comparison} demonstrates that the max-min multi-commodity flow method already outperforms a standard shortest-path finding algorithm, as it allows for the large amount of keys to be split up and send over multiple links. The algorithm becomes even more interesting when actually handling multiple key exchanges at once. Whereas in case of the Dijkstra algorithm, the order of the requests is relevant, the Max-Min algorithm treats all requests equally, ensuring that all GS pairs can exchange a reasonable amount of keys.  


\begin{table}[t]
    \centering
    \caption{Comparison of elementary results from different methods. Multiple requests interpreted as sequential in Dijkstra.  
    }
    \vspace{-2mm}
    \label{tab: simple_comparison}
    \begin{tabular}{|c|*{4}{c|}}
        \hline
    \rowcolor{light_grey}  \backslashbox{Path}{Method} & Max-min demand  & (sequential) Dijkstra \\ 
     \hline
     $A \mapsto B$ & 27,000 & 24,000 \\
     \hline
     $A \mapsto E$ & 3,600 & 600 \\
     \hline
     $C \mapsto D$ & 3,600 & 600\\
     \hline
     $C \mapsto A$, $B \mapsto A$ & 13,500 \& 13,500 & 24,000 \& 2,400\\ 
     \hline
     $B \mapsto A$, $C \mapsto A$ & 13,500 \& 13,500 & 24,000 \& 2,400\\ 
     \hline
    \end{tabular}
          \vspace{-3mm}
\end{table}


With five GSs as in Fig.~\ref{fig: simulation_graph}, there are $\binom{5}{2} = 10$ unique source sink combinations up to direction. Algorithm~\ref{alg: max_min} finds out that it is possible to send at most $600$ secret keys bits per GS pair before the QKD network becomes in outage. Note that this does not mean that no more additional secret key could be sent. Actually, when running Algorithm \ref{alg: min_resource} with requests of size $600$, it is possible to see that most of the links of the QKD network still have quantum keys in its pools. The only links in outage are those connecting the group $\{A,B,C\}$ to $\{D ,E\}$. This is because any exchange of secret keys between a GS from $\{A,B,C\}$ to a GS in $\{D,E\}$ must go through an inter-satellite link, which has a much lower secret key generation rate than a LEO-to-GS links due to the difficulty of placing large payloads on the space. This effect is well illustrated by the performance of the Max-Min multi-commodity flow algorithm, in Table~\ref{tab: grouping}, which shows the minimal demands filled by Algorithm~\ref{alg: max_min} on few sample combinations. This table summarizes the amount of secret keys that would be consumed by sharing the given amount of keys on each pair of GSs, according to Algorithm \ref{alg: min_resource}, and the corresponding consumption rate, which is the amount of used keys per successfully sent out key. Indeed for the combinations involving GS~$A$, GS~$B$, and GS~$C$,the algorithm finds that a minimum demand of $13,500$\,bits is feasible.

\vspace{-2mm}
\section{Conclusion and Future Extensions}
\label{sec:5}
\vspace{-0.5mm}

This paper studied the resource allocation problem of a QKD network that combines both GEO and LEO quantum satellites, acting as trusted-repeaters, which use the BB84 protocol with decoy state to generate secret keys that are consumed when GS pairs exchange encryption keys securely. Starting from the graph representation of the QKD network in a given time window, an equivalent LP problem was presented. Then, algorithms were derived to maximize the minimum amount of keys exchanged in each GS pair, and to minimize the amount of secret keys consumed in the space-to-ground and inter-satellite links to fulfill a given demand. 

The results suggest that the max-min multi-commodity flow algorithm exhausts the link between two subsets of the nodes of the graph, with plenty of resources on links within the subsets. This could be taken into account by using the max-min flow algorithm recursively on the subsets until no more subsets can be created to fully exhaust the network. 

The assumption to allow a GS to act as a trusted-repeater may be questionable. A straightforward solution is to allow only one link in each GS. A more complex solution, which would allow multiple connections, is to take out the GS nodes from the flow graph, give each commodity multiple sinks and sources, and link the GEO/LEO satellites with the GS. 
The restrictions of the demands would be based on the size of the quantum key pools of the links. Similarly, a linear system of the flows and demands can be created, to be solved using LP. 

\begin{table}[t]
    \centering
    \caption{Results for the requests handled by GS combinations.} 
    \vspace{-2mm}
    \label{tab: grouping}
        \begin{tabular}{|c|c|c|c|}
    \hline
        \rowcolor{light_grey} Combi- & \hspace{-2.5mm} Min. filled demand \hspace{-2.5mm}  & \hspace{-2.5mm} Total consumed \hspace{-2.5mm} & \hspace{-5mm} Consumption rate \hspace{-5mm}  \\
         \rowcolor{light_grey} -nations & by MMD & \hspace{-2.5mm} keys by MR \hspace{-2.5mm} & \hspace{-2.5mm} (\#used keys/\#sent key) \hspace{-5mm}\\
        \hline
        A, B & 27,000 & 56,400 & 2.09 \\
        \hline
        A, D & 3,600 & 19,200 & 5.3 \\
        \hline
        A, B, C & 13,500 & 106,800 & 2.64 \\
        \hline
        A, B, D & 1,800 & 20,400 & 3.78 \\ \hline
        \hspace{-2.5mm} A, B, D, E \hspace{-2.5mm} & 900 & 18,000 & 3.33\\
        \hline
        All GSs & 600 & 18,000 & 3.00 \\ \hline
    \end{tabular}
      \vspace{-5mm}
\end{table}

\vspace{-2mm}
\section*{Acknowledgment}

This publication has been based upon work from COST Action CA19111 NEWFOCUS, supported by COST (European Cooperation in Science and Technology).

\vspace{-2mm}

\bibliographystyle{IEEEtran}
\bibliography{Transactions-Bibliography/IEEEabrv, Transactions-Bibliography/references}



\end{document}